\documentclass[]{emulateapj}

\slugcomment{version 20090225, identical to the published one}

\shorttitle{Lyman Continuum from z=3 Galaxies}
\shortauthors{Iwata et al.}

\begin{document}


\title{Detections of Lyman Continuum from Star-forming Galaxies at $z\sim3$ 
 Through Subaru/Suprime-Cam Narrow-band Imaging\altaffilmark{1,2}}
 

\author{I. Iwata\altaffilmark{3}, A. K. Inoue\altaffilmark{4},
Y. Matsuda\altaffilmark{5,6},  H. Furusawa\altaffilmark{7}, 
T. Hayashino\altaffilmark{8}, K. Kousai\altaffilmark{8}, 
M. Akiyama\altaffilmark{7,9}, T. Yamada\altaffilmark{9}, 
D. Burgarella\altaffilmark{10},  and J.-M. Deharveng\altaffilmark{10}} 

\altaffiltext{1}{Based on data collected with the Subaru Telescope,
which is operated by the National Astronomical Observatory of Japan.}

\altaffiltext{2}{Some of the data presented in this paper were obtained
from the Multimission Archive at the Space Telescope Science Institute
(MAST). STScI is operated by the Association of Universities for
Research in Astronomy, Inc., under NASA contract NAS5-26555. 
}

\altaffiltext{3}{Okayama Astrophysical Observatory, National
    Astronomical Observatory of Japan, Honjo, Kamogata, Asakuchi, 
    Okayama 719-0232; iwata@oao.nao.ac.jp} 

\altaffiltext{4}{College of General Education, Osaka Sangyo University, 
3-1-1 Nakagaito, Daito, Okaka 574-8530, Japan; akinoue@las.osaka-sandai.ac.jp}

\altaffiltext{5}{Department of Astronomy, Graduate School of Science, 
Kyoto University, Kyoto 606-8502, Japan}

\altaffiltext{6}{Optical and Infrared Astronomy Division, 
National Astronomical Observatory of Japan, 
2-21-1 Osawa, Mitaka, Tokyo 181-8588; yuichi.matsuda@nao.ac.jp}

\altaffiltext{7}{Subaru Telescope, National Astronomical Observatory of
Japan, 650 North A'ohoku Place, Hilo, HI 96720.; furusawa@subaru.naoj.org}

\altaffiltext{8}{Research Center for Neutrino Science, Graduate School
of Science, Tohoku University, Sendai 980-8578, Japan;
haya@awa.tohoku.ac.jp, kousai@awa.tohoku.ac.jp} 

\altaffiltext{9}{Astronomical Institute, Graduate School of Science,
Tohoku University, Aramaki, Aoba-ku, Sendai 980-8578, Japan;
akiyama@astr.tohoku.ac.jp, yamada@astr.tohoku.ac.jp}

\altaffiltext{10}{Laboratoire d'Astrophysique de Marseille, UMR 6110,
CNRS/Universite de Provence, 38 rue Joliot-Curie, 13388 Marseille Cedex
13, France; 
Denis.Burgarella@oamp.fr, jean-michel.deharveng@oamp.fr}



\begin{abstract}
 Knowing the amount of ionizing photons from young star-forming
 galaxies is of particular importance to understanding the reionization
 process. Here we report initial results of Subaru/Suprime-Cam deep
 imaging observation of the SSA22 proto-cluster region at $z=3.09$,
 using a special narrow-band filter to optimally trace ionizing
 radiation from galaxies at $z\sim3$. The unique wide field-of-view of
 Suprime-Cam enabled us to search for ionizing photons from 198 galaxies
 (73 Lyman break galaxies (LBGs) and 125 Ly-$\alpha$ emitters (LAEs))
 with spectroscopically measured redshifts $z \simeq 3.1$. We detected
 ionizing radiation from 7 LBGs, as well as from 10 LAE candidates. 
 Some of the detected galaxies show significant spatial offsets of
 ionizing radiation from non-ionizing UV emission. 
 For some LBGs the observed non-ionizing UV to Lyman continuum flux
 density ratios are smaller than values expected from population
 synthesis models with a standard Salpeter initial mass function (IMF)
 with moderate dust attenuation (which is suggested from the observed UV 
 slopes), even if we assume very transparent IGM along the sightlines of
 these objects. 
 This implies an intrinsically bluer spectral energy distribution, e.g,
 that produced by a top-heavy IMF, for these LBGs.
 The observed flux desity ratios of non-ionizing UV to ionizing radiation
 of 7 detected LBGs range from 2.4 to 23.8 and the median is 6.6.
 The observed flux density ratios of the detected LAEs are even smaller 
 than LBGs, if they are truly at $z \simeq 3.1$. 
 We find that the median value of the flux density ratio for the deteced
 LBGs suggest that their escape fractions is likely to be higher than
 4\%, if the Lyman continuum escape is isotropic. The results imply that
 some of the LBGs in the proto-cluster at $z\sim3$ have the escape
 fraction significantly higher than that of galaxies (in a general
 field) at $z\sim1$ studied previously.
\end{abstract}


\keywords{galaxies: evolution --- galaxies: high-redshift --- 
cosmology: observations --- intergalactic medium --- diffuse radiation} 



\section{Introduction}

Ionizing radiation from star-forming galaxies is
a likely primary source of cosmic reionization.
Although the ratio of the flux density of Lyman continuum escaping 
from a galaxy to that produced in the galaxy, the so-called escape
fraction ($f_\mathrm{esc}$), is a key parameter for evaluating the
contribution of galaxies to cosmic reionization, it has been poorly
constrained due to the fact that Lyman continuum photons are easily absorbed by
the intergalactic medium (IGM).
Direct observations of Lyman continuum from galaxies at $z>4$ are
virtually impossible because of a rapid increase in the number density
of Lyman limit systems toward high redshifts \citep{inoue08a}. 
Therefore, 
we must focus on $z\sim3$ where the IGM optical depth is still about unity on
average. 
\citet{steidel01} detected Lyman continuum in a composite spectrum
of 29 Lyman break galaxies (LBGs) at $\langle z\rangle=3.4$. 
The observed flux density ratio (i.e., without correction for IGM
absorption) was $f_{1500}/f_{900}=17.7\pm3.8$. 
\citet[][hereafter S06]{shapley06} detected Lyman continuum
individually from two out of 14 observed LBGs at $z\sim3$ 
through a deep long-slit spectroscopy, and derived 
$f_{1500}/f_{900}=12.7\pm1.8$ and $7.5\pm1.0$ for the two LBGs. 
For an average of 14 objects, $f_{1500}/f_{900}=58\pm35$. 
So far the number of galaxies with direct observation of Lyman
continuum is too small to reveal properties shared by galaxies 
with large $f_\mathrm{esc}$, or to estimate a typical value of 
$f_\mathrm{esc}$.

In this paper we report the initial results of the observations
of Lyman continuum from galaxies at $z \sim 3$ with Subaru/Suprime-Cam
\citep{miyazaki02}. 
A narrow-band filter imaging was adopted to search for Lyman continuum,
as pioneered by \citet{inoue05} for galaxies at $z\sim 3$. The use of
narrow-band imaging instead of slit spectroscopy enables us to examine
ionizing radiation from a large number of galaxies simultaneously, as well as
to examine emission offsets from rest-frame non-ionizing UV radiation. 
The target field of the present study is the SSA22 field where the 
prominent proto-cluster of galaxies at $z=3.09$ has been discovered 
\citep[][hereafter H04]{steidel98, steidel00, hayashino04}.
We adopt a cosmology with the parameters $\Omega_\mathrm{M}=0.3$,
$\Omega_\Lambda=0.7$ and $\mathrm{H_0}=70$ km/s/Mpc. 
All magnitudes are measured in the AB system.

\section{Observations and Data Reduction}

\begin{figure}
\plotone{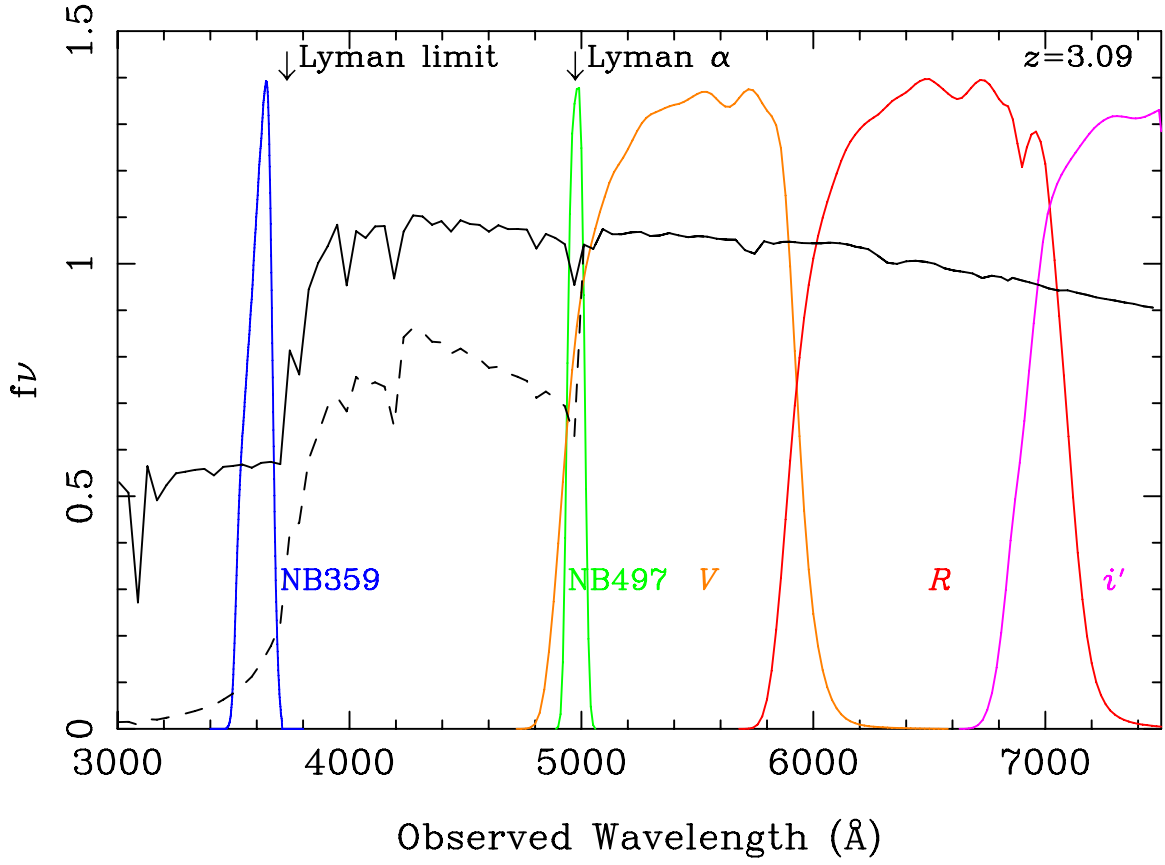}
 \caption{Filter transmission curves for the filters used in this study
 and spectra of a model star-forming galaxy at $z=3.09$ generated with
 the Starburst99 code. The solid line represents a model without IGM
 attenuation, and the dashed line is for a model with average IGM
 attenuation for objects at $z=3$. See section 4.3 for details of the
 model.
\label{fig:sed}}
\end{figure}

We have built a special narrow-band filter NB359 with a central
wavelength of 359 nm and a FWHM of 15 nm (and $>10$\% transmittance
between 350 nm and 371 nm). This filter is designed to trace Lyman
continuum of galaxies at $z\gtrsim 3.06$ with a contamination from
non-ionizing radiation less than 1\% for a typical star-forming
galaxy.
Laboratory measurements have shown that the transmission of
the filter is less than 0.01\% in the wavelength range of 400 nm--1200 nm. 
We also confirmed that the central wavelength is stable with rms=0.47
nm at every position on the filter.
In Figure \ref{fig:sed} we show the transmission curve of the NB359
filter as well as transmission curves of NB479 which was used in the
survey for Ly-$\alpha$ emitters (LAEs) at $z=3.1$ (H04), $V$, $R$,
$i'$ band filters and spectra of model star-forming galaxies at
$z=3.09$, which is used in section 4.3. 

The imaging observations using Subaru/Suprime-Cam with the NB359 filter 
were carried out in 2007 September 10--14 (UT).
Three nights were photometric while in the other two nights cloud
coverage was relatively high. The seeing condition was good throughout
the observing dates, with a FWHM of $0.''5$--$0.''9$ in NB359 images. 
The pointing center 
($\alpha=22^\mathrm{h}17^\mathrm{m}26^\mathrm{s}.1$, 
$\delta=+00\degr16\arcmin11\arcsec$ [J2000]) 
is aligned to the previous Subaru/Suprime-Cam observations of the
SSA22 field (Matsuda et al.~2004, hereafter M04; H04).
For half of the shots the position angle was switched from 90\degr to
270\degr, in order to reduce the effect of chip-to-chip sensitivity
variation.

The data reduction was made using SDFRED
version 1.2.5 \citep{yagi02,ouchi04}. 
The rejection of frames under poor sky condition led to a selection of
45 frames with 1,800 sec exposures to produce the mosaic image.
The FWHM of point sources is 0.\arcsec8. 
The flux calibration was made using measurements of spectroscopic
standard stars. 
The 3 $\sigma$ limiting magnitude of the final mosaic image is
estimated to be 27.33 AB mag for $1.''2$ diameter aperture. 
Source detection was done using SExtractor \citep{bertin96} version
2.5.0, using both $R$-band and NB359 images.
Galactic extinction has been corrected using the dust map by 
\citet{schlegel98}. 

\section{Results}

There are 198 objects with spectroscopically measured redshifts larger
than 3.0 in the Suprime-Cam field-of-view ($32'\times 26'$). Ten objects 
classified as QSOs/AGNs from optical spectra and/or X-ray observation
have already been excluded. 
Among the 198 objects there are 44 LBGs reported in the literature 
\citep[][S06]{steidel03}, 29 LBGs selected as
$U$-dropout galaxies and identified spectroscopically with VLT/VIMOS
(Kousai et al., in preparation), and 125 Ly-$\alpha$ emitters (LAEs) and
Ly-$\alpha$ 'blobs' selected through a narrow-band (NB497) imaging with
Subaru/Suprime-Cam (H04, M04) 
and identified spectroscopically with Subaru/FOCAS and Keck/DEIMOS 
\citep[][Matsuda et al., in preparation]{matsuda06}. 

\begin{figure}
\plotone{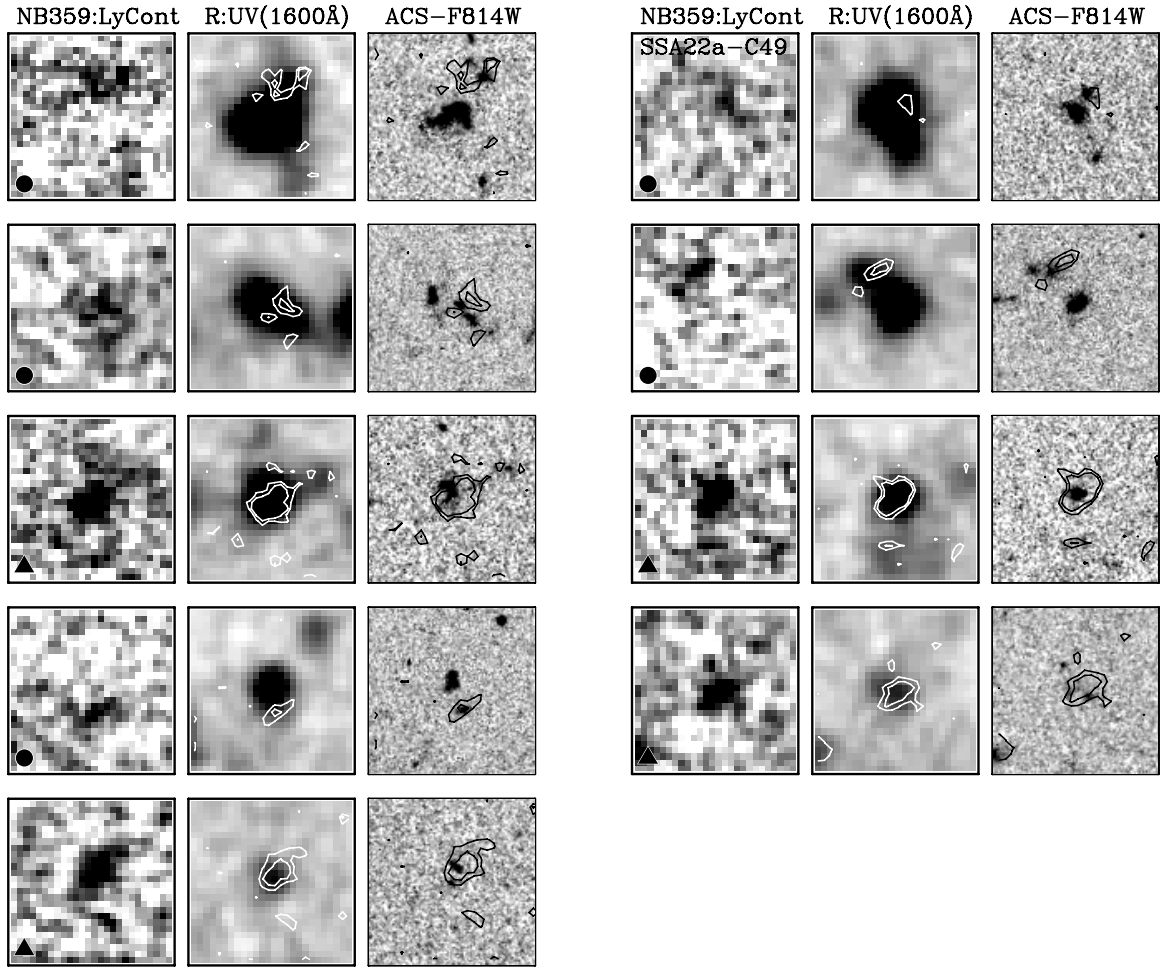}
 \caption{Postage stamp images of objects which are detected in the 
 NB359 image and HST/ACS F814W images are available for them. 
 In each panel NB359 (left), $R$-band (middle) and F814W (right) images
 are shown.  Field of view is 5$''$ $\times$ 5$''$. 
 For the $R$-band and F814W images a contour map of the NB359 image 
 (2 and 3 $\sigma$) is over-plotted. 
 Symbols at the lower-left of the NB359 images show the object type:
 filled circles: LBG, filled triangle: 'blue' LAEs. See section 4.3 for the
 definition of types.
 The images of SSA22a-C49, which is one of the two objects repoted to
 be detected in S06 and is detected in our NB359 image with 2.95$\sigma$
 level, is labelled.
\label{fig:montage1}}
\end{figure}

\begin{figure}
\plotone{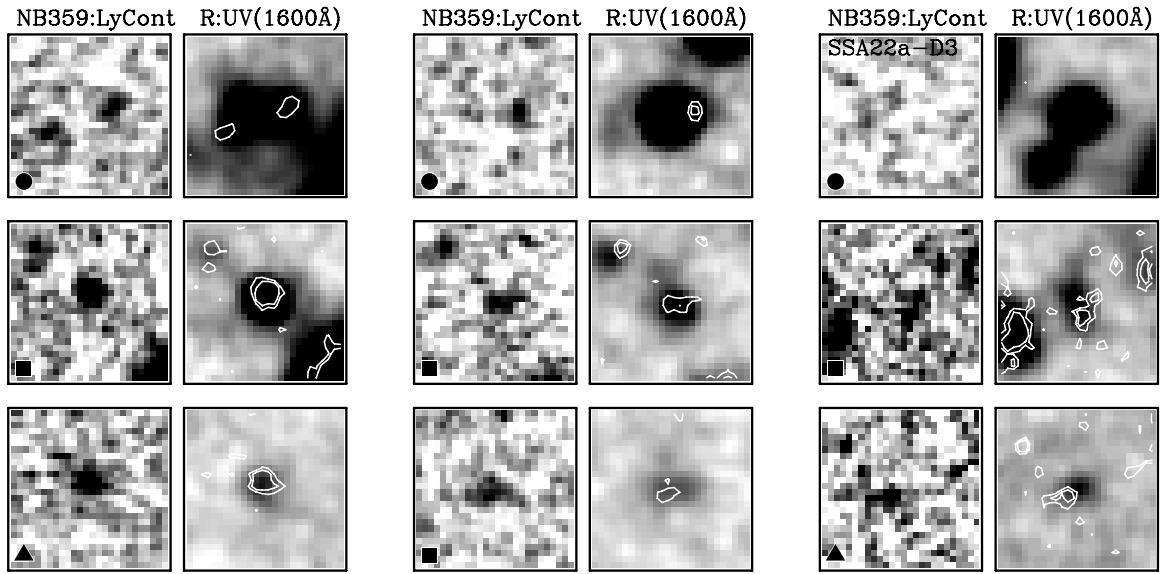}
 \caption{Postage stamp images of objects which are detected in the
 NB359 image (HST/ACS F814W images are not available). 
 In each panel NB359 (left) and $R$-band (right) images
 are shown.  Field of view is 5$''$ $\times$ 5$''$. 
 For the $R$-band images a contour map of the NB359 image 
 (2 and 3 $\sigma$) is over-plotted. 
 Here we also show images of SSA22a-D3, which is one of the two objects
 reported to be detected in \citet{shapley06} but is not detected in our
 NB359 image.
 Symbols at the lower-left of the NB359 image show the object type:
 filled circle: LBG, filled triangle: 'blue' LAEs, and filled squares:
 'red' LAEs. See section 4.3 for the definition of types.
\label{fig:montage2}}
\end{figure}

We detected 16 objects with secure detection ($>3\sigma$) within 
1.\arcsec2 diameter apertures in NB359;
6 are LBGs and 10 are LAEs. The redshift range of the LBGs is
$3.04<z<3.31$, while for the LAEs it is $3.07<z<3.10$.
In order to eliminate a possibility of spurious detections, we split 
the frames used to create the final NB359 image into two sets and
generated two mosaic images. All 16 objects were detected in both
images at $\gtrsim2\sigma$ level. 
We also executed a detection test using a negative image in NB359 and
found that a probability of spurious detection at $>3\sigma$ level 
is $\sim0.4$\%. 
Therefore, we are confident in the reality of source detection in NB359.
In Figures \ref{fig:montage1} and \ref{fig:montage2} we show 
$5''$ $\times$ $5''$ postage stamp images of the detected objects. The
objects shown in Figure \ref{fig:montage1} are within fields observerd 
with the Advanced Camera for Surveys (ACS) on board the Hubble Space
Telescope \citep[HST;][]{abraham07, geach07}, and their F814W images are
shown in the figure.

In addition, the object SSA22a-C49, detected in Lyman continuum by S06,
has 2.95$\sigma$ significance in our image. 
The measured flux density within the 1.\arcsec2 aperture 
($5.5\pm1.0\times10^{-2}$ $\mu$Jy) is consistent with that measured with 
a long-slit spectroscopy by S06 
($6.9\pm1.0\times10^{-2}$ $\mu$Jy). 
We add this object to the detection list in NB359, bringing the
total number of detected objects to 17 (7 LBGs and 10 LAEs). 
The images of SSA22a-C49 are shown in Figure \ref{fig:montage1}. 
On the other hand, we could not find any significant signal in NB359 at
the position of SSA22a-D3 which is another object detected in Lyman 
continuum by S06. 
Since the 3$\sigma$ detection limit within 1.\arcsec2 aperture at the
position of this object ($5.8\times 10^{-2}$ $\mu$Jy) is well below the
flux density reported by S06 
($11.8\pm1.1\times10^{-2}$ $\mu$Jy), it is not clear why the
object is not visible in our NB359 image.
The images of SSA22a-D3 are also presented in Figure \ref{fig:montage2}.

All of the 10 LAEs detected in NB359 show a prominent emission line
around 497 nm in their spectra taken with FOCAS and DEIMOS, but continua
are not well detected due to their faintness. 
All but one spectrum of these 10 objects have a resolution high 
enough to distinguish the doublet if the emission line detected with
the NB497 filter is [O {\sc ii}] $\lambda$3727 at $z=0.33$, and we find no 
such feature. However, there is still a possibility that they are AGNs
at lower redshifts if the emission line is C {\sc iv} $\lambda$1549 at
$z=2.21$, [C {\sc iii}] $\lambda 1909$ at $z=1.60$ or Mg {\sc ii}
$\lambda$2798 at $z=0.78$, although some of them 
show a spatially extended emission in the NB497 image. 
Unfortunately, the wavelength coverage of the spectra obtained so far is
too narrow to definitely rule out the low redshift
possibility. Follow-up spectroscopy would be required to clarify 
whether these objects are really LAEs at $z\simeq3.09$. 

If a faint foreground object lies very close to an object at 
$z \simeq 3.1$, they might mimic Lyman continuum in our NB359, and
it would be difficult to distinguish it at longer wavelengths if the
object at $z \simeq 3.1$ is brighter than the foreground object. 
\cite{siana07} discussed such a possibility of contamination by faint
foreground objects in the $z\sim3$ Lyman continuum survey. Following
\citet{siana07}, we roughly estimate the probability of such a case by
assuming a surface density of faint galaxies and their spatial
distribution to be uncorrelated with the $z\sim3$ objects.
The apparent magnitudes of galaxies detected in NB359 range from 26.5
mag to 27.5 mag. The surface number density of galaxies in this
magnitude range from $U$-band number count by \citet{williams96} is
$\sim10^5$ mag$^{-1}$ deg$^{-2}$. If we consider that a foreground
object within a 1.\arcsec0 radius cannot be distinguished in our NB359
image, each object has $\sim$2.4\% chance of such foreground
contamination. Since we have 198 spectroscopic sample galaxies, 
several objects among the galaxies detected in NB359 may be explained
by contamination by such a foreground object. 
However, it would be difficult to imagine that all
17 detections could be due to foreground contamination.

\section{Discussion}

\subsection{Spatial Offsets of Lyman Continuum for LBGs}

As seen in Figures \ref{fig:montage1} and \ref{fig:montage2}, shapes and
positions of the emitting regions in the 17 detected galaxies in NB359
are quite different from those in the $R$-band, especially for LBGs
(marked with circles in Figures \ref{fig:montage1} and
\ref{fig:montage2}). 
The rms of positional offsets between the NB359 and the $R$-band images
for foreground objects with similar magnitude range ($23.5<R<27.0$) 
is $\sim0.\arcsec25$. 
The offsets of 4 LBGs among the 7 LBGs detected in Lyman continuum 
exceed $3\sigma$, and the average offset is 0.\arcsec97 
(3.8$\sigma$, corresponding to 7.4 kpc at $z=3.09$). 
Higher resolution of HST/ACS images clearly shows that in each object
with positional offset between NB359 and $R$-band images there are
substructures in F814W which agree with the positions of the regions
detected in NB359.
Except for a few possible cases of foreground contaminations 
(as discussed in section 3),
such differences in the shape and the position between the emitting
regions of ionizing radiation and those of non-ionizing UV may give us a
clue to understand how Lyman continuum escapes from galaxies. 
For instance, Lyman continuum may escape through a chimney-like
structure in the interstellar medium \citep[][]{razoumov07}, and we may
see the emission only from some limited regions. 
Another possibility is that the spatial distribution of 
the Lyman continuum sources is different from that of the non-ionizing
UV emitting stars. 
As shown in Figure \ref{fig:montage1}, the higher spatial resolution of
HST/ACS reveals that many of the detected galaxies have multiple knotty
structures and that the Lyman continuum emission appears to be
associated with one of such knots.
Note that 
at the moment with the F814W image we cannot 
distinguish between superposition of a foreground object and multiple 
knotty substructures frequently seen in LBGs \citep[e.g.,][]{lowenthal97}. 
The position offsets of Lyman continuum from non-ionizing UV peak
also imply a possibility that long-slit spectroscopy 
may miss Lyman continuum.

\subsection{Photometry and Colors of the Detected Objects}

Because of the various spatial offsets in the Lyman continuum, using a
fixed aperture size to measure the flux for all detected sources is not
appropriate to examine the flux ratio between the Lyman continuum and
non-ionizing UV emission from these systems. 
In order to obtain total flux in each image we put the apertures with
different radii centered at the peak of an object in $R$-band, examined
the curves of growth in NB359 and $R$-band and determined the aperture
sizes which are large enough to contain the bulk of fluxes in both
images and which do not contain flux from nearby objects. The aperture
diameters for the 17 detected galaxies are $2.''0$--$4.''0$.
In Figure \ref{fig:color1} the NB359$-R$ colors of the 198 objects with
spectroscopic redshifts larger than 3.0 are plotted 
against their $R$-band magnitudes. The NB359$-R$ color corresponds to 
the apparent flux density ratio of ionizing to non-ionizing UV photons
for objects at $z\sim3$.
For objects detected in the NB359 image, variable aperture sizes
described above are used to measure colors, while using arrows we also
show colors measured with a $1.''2$ aperture, which was used in object
detection, both in NB359 and $R$-band. 
We find that luminous ($R\lesssim25$) objects, which roughly correspond 
to those with $L>L^\ast$ of LBGs at $z\sim3$ \citep{sawicki06}, show
relatively red colors of NB359$-R \gtrsim 2$. On the other hand, 
some of the less luminous objects show bluer colors, 
$0 \lesssim \mathrm{NB359}-R < 2$. 
If the IGM opacity does not depend on the source luminosity, this
suggests that the ionizing-to-non-ionizing UV escape flux density ratio 
(a proxy of $f_\mathrm{esc}$; see \citealt{inoue06}) is
lower in relatively luminous objects. Interestingly, this appears 
contrary to the argument by \cite{gnedin08b} that $f_\mathrm{esc}$ is
smaller for galaxies with smaller star formation rates (or mass).
However, we should note that, as seen in Figure \ref{fig:montage1},
Lyman continuum appear to be emitted from substructures in UV-luminous
objects, and the NB359$-R$ colors of such substructures would be much
bluer than the colors using total flux densities.

\subsection{Comparison with Model Stellar Populations}

We use deep multi-band optical imaging data (H04) 
to investigate spectral energy distribution (SED) of the detected sample. 
The UV spectral slopes of the detected LBGs are flat ($0\lesssim V-i'<0.35$). 
There are two clearly distinctive sub-groups of the detected LAEs from 
their $V-i'$ colors: one is red ($0.4<V-i'\lesssim 0.8$), and the other is 
blue ($-0.5<V-i'<0$).
In Figure \ref{fig:color1} different symbols are used for different types of
SEDs. Blue LAEs show extremely blue NB359$-R$ colors
($<0.5$), while LBGs have NB359$-R$ larger than 0.9. 
The difference between these three types becomes clear if we place them 
into the NB359$-R$ and $V-i'$ two-color plane, as shown in Figure
\ref{fig:color2}. 
To compare with these observed colors 
we calculated predictions of colors for young
star-forming galaxies 
with a population synthesis code
\citep[Starburst99 version 5.1;][]{leitherer99}.
A model at zero age, with the Salpeter IMF 
\citep[0.1--120$M_\sun$;][]{salpeter55}  
and Padova evolutionary tracks 
with metallicity $Z=4\times10^{-4}$ is used as a fiducial one. 
Since models with higher metallicity or older age have redder colors, 
this model has the bluest SED under an assumption of the Salpeter IMF. 
The colors of the model without IGM and dust attenuation and assuming
$f_\mathrm{esc}=1$ at $z=3.0$ and 3.3 are shown as filled and open
circles connected with a solid line in Figure \ref{fig:color2}. 
In Figure \ref{fig:color2}(a) the model does not include flux from
nebular continuum and only flux from stellar sources is considered. 
An arrow indicates the direction of dust
attenuation following a prescription by \citet{calzetti00} for a galaxy at
$z=3$. For $\lambda < 1200$\AA\ we simply extrapolated their attenuation
law. Such smooth extrapolation toward extreme-UV may be reasonable up to
$\lambda \simeq 800$ \AA\ \citep{draine03}.
IGM attenuation has a large dispersion for different lines of sight. 
We show colors with a median value of IGM attenuation for $z=3.0$ and
$z=3.2$ cases, with the optical depth range containing 68\% of all
sightlines of Monte-Carlo simulations \citep{inoue08a}. 
The shaded area shows colors of model galaxies which can be explained
with dust and/or IGM attenuation. The dashed line in the area indicates
colors with IGM opacity with 1\% probability of occurrence 
(i.e., very rare transparent sight-lines). 

The $V-i'$ colors of the detected LBGs suggest moderate attenuation by
dust,  $E(B-V) \sim 0.1$--0.3, consistent with previous studies 
\citep[e.g.,][]{iwata05}. However, the NB359$-R$ colors of the detected
LBGs are difficult to explain with these models. 
Even if we assume no IGM attenuation, 
NB359$-R$ colors of the three bluest LBGs cannot be explained. 
One caveat is that the model SEDs considered in Figure
\ref{fig:color2}(a) do not include nebular continuum. The output SEDs by
Starburst99 code have nebular continuum emission assuming all stellar
ionizing photons are converted to nebular continuum photons. In Figure
\ref{fig:color2}(b) we show colors of the model with both stellar and
nebular continua (i.e., the model is not self-consistent in regard to
photon budget). The inclusion of nebular continuum makes SEDs redder in
both NB359$-R$ and $V-i'$, and in this case the colors of the four
redder LBGs would be explained with moderate IGM attenuation.
However, there are still three LBGs whose colors cannot be explained 
with the bluest SED and Calzetti's dust attenuation law. 
A QSO-like spectrum may be able to explain these colors. 
The power-law slope of QSO UV spectrum have large variation. Here we
assume two-component power law ($F_\nu \propto \nu^\alpha$) with a break
at 1100\AA, as it is made in the appendix of \citet{inoue08a}, and
consider the possible bluest color of a QSO at $z\sim3$.
The power law index at $\lambda < 1100\mathrm{\AA}$ can be as steep as
$-0.6$ \citep{scott04}. At $\lambda > 1100\mathrm{\AA}$ the steepest
slope would be $\sim -1.8$ \citep{shang05}. In such a case NB359$-R$ and 
$V-i'$ are $\sim0.23$ and $0.07$, respectively, without IGM
attenuation. Thus by considering IGM attenuation such QSO-like spectrum
would be able to explain the colors of the LBGs shown in Figure
\ref{fig:color2}, although these objects are not
classified as AGNs, according to the catalog \citep{steidel03}. 

To explain these strong emission in Lyman continuum without considering
possibility of QSOs, models which 
can produce much bluer intrinsic colors -- such as those with a
top-heavy IMF -- would be required. Indeed, model SEDs with a top-heavy 
IMF with a slope 1.0 
(much steeper than the slope of the Salpeter IMF, 2.35) have  
$\sim 0.3$ mag bluer color in NB359$-R$. 
We also examined colors of the SEDs with zero
metallicity stellar population (Population III) with the Salpeter IMF
(mass range from 1$M_\sun$ to 100$M_\sun$) by
\citet{schaerer03}. In Figure \ref{fig:color2}(a) and (b) the colors
of the model at $z=3.0$--$3.3$ without and with nebular continuum are
plotted, respectively. If no nebular continuum is included (Figure
\ref{fig:color2}(a)), the NB359-$R$ colors of the Pop-III models are 
$\sim0.3$ mag bluer than those of Starburst99 $Z=4\times10^{-4}$ models
(with Salpeter IMF) and are close to the top-heavy cases. 

Alternatively, dust attenuation law might be quite different from
that by \citet{calzetti00}. For example, if the amount of dust
attenuation at 900\AA\ is much smaller than that at 1500\AA, the dust
attenuation arrow approaches vertical in Figure \ref{fig:color2}, and 
the NB359$-R$ and $V-i'$ colors of the LBGs may be explained
simultaneously. 

Colors of the detected LAEs are much harder to explain with the models
considered here. 
If higher mass cut-off (50$M_\odot$--500$M_\odot$) is adopted,
NB359-$R$ color of the Population III model by \citet{schaerer03} 
(using Salpeter IMF and no IGM attenuation is considered) is as
small as $-0.5$, and it would be compatible with the bluest LAEs. Thus
we would like to stress that, although some of the NB359$-R$ colors of
the detected objects are unexpectedly blue, such colors are not
physically unexplainable and they can be reproduced if we consider
metal-poor stellar populations and/or top-heavy IMF. 
Note that there is a suggestion of the existence of metal-free stellar
populations at $z\approx3$ based on observed data \citep{jimenez06}.
For a reference, there are three [O {\sc ii}] emitters at $z\sim0.33$
found by the follow-up spectroscopy of the SSA22 LAE survey (Matsuda et
al., in preparation). Their colors in NB359$-R$ and $V-i'$ are
0.9$\pm$0.3 and 0.1$\pm$0.1, respectively, and they are similar to some
of LBGs with Lyman continuum detections and are different from those of
the detected LAEs. 
Anyway, deep spectroscopy with wide wavelength coverage is required to
confirm their redshifts and explore their nature. 
In the following discussion for constraints on $f_\mathrm{esc}$, 
we only use the detected LBG sample.

\subsection{Constraints on the Escape Fraction}

\begin{figure}
\plotone{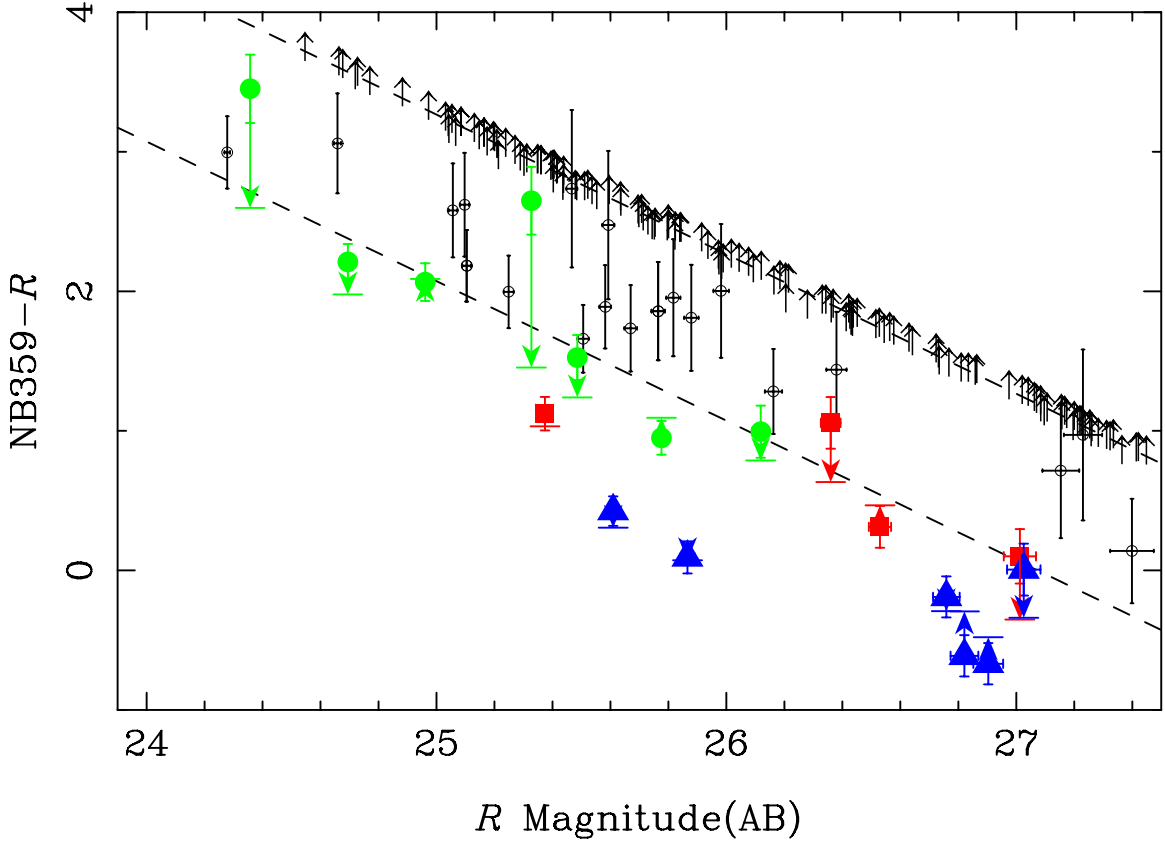}
\caption{NB359$-R$ colors of objects in the SSA22 field with
 spectroscopic redshifts $z>3.0$. The higher and lower dashed lines
 indicate approximately $1\sigma$ and $3\sigma$ limits in the NB359
 image. The objects without significant detection ($\lesssim 3\sigma$)
 in the NB359 image are plotted with open circles and upper arrows. 
 The detected LBGs are shown with filled circles.
 LAEs are classified with two types according to their rest-frame UV
 slopes: red UV colors (filled squares) and blue UV colors (triangles). 
 See text for details. 
 For the objects detected in the NB359 image, 
 NB359$-R$ colors are measured with variable aperture sizes centered at
 peak positions in $R$-band determined to encompass total fluxes, 
 and arrows indicate how colors change if 
 $1.''2$ apertures centered at peaks in NB359 and $R$-band are adopted.
\label{fig:color1}}
\end{figure}

\begin{figure}
\plottwo{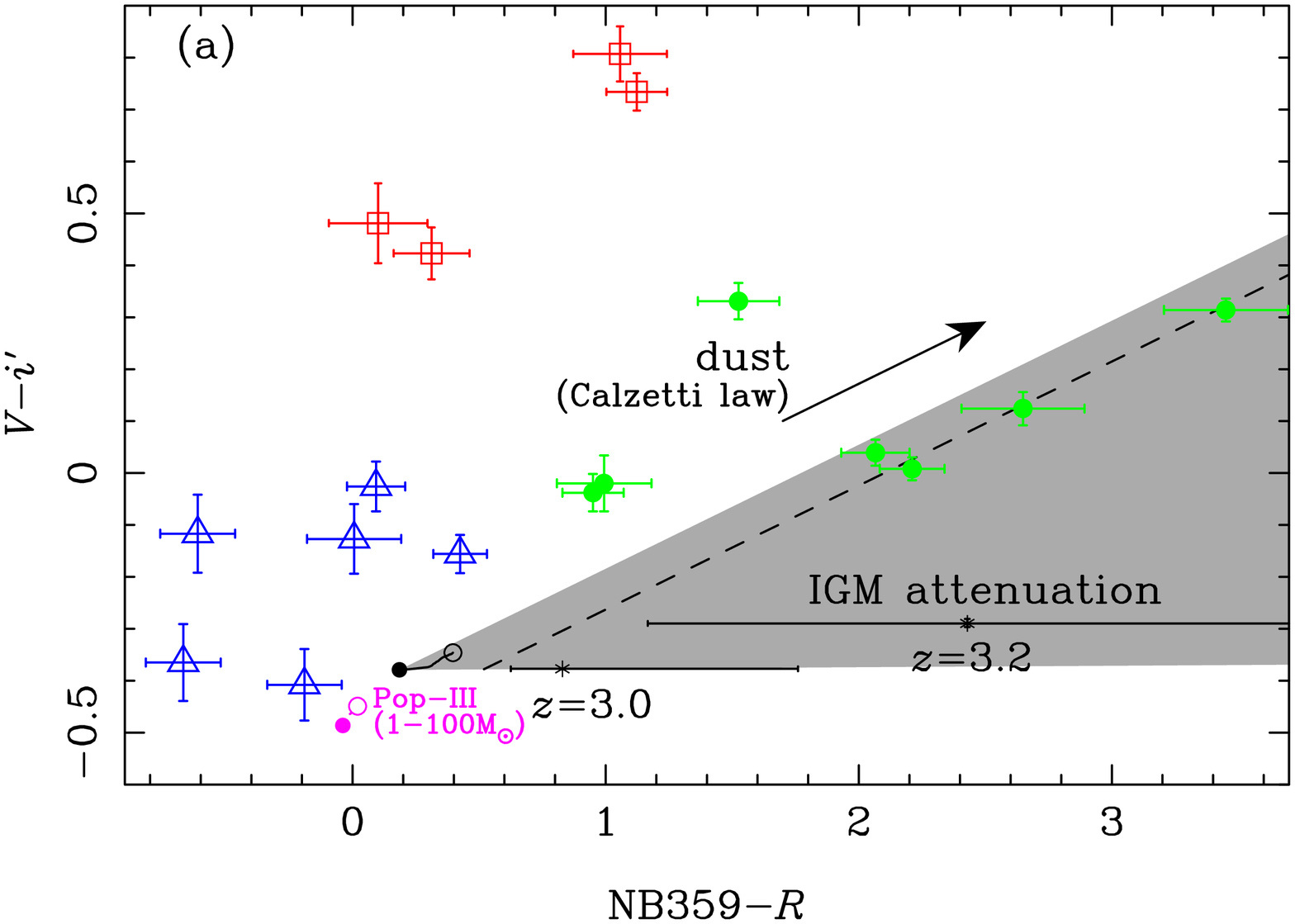}{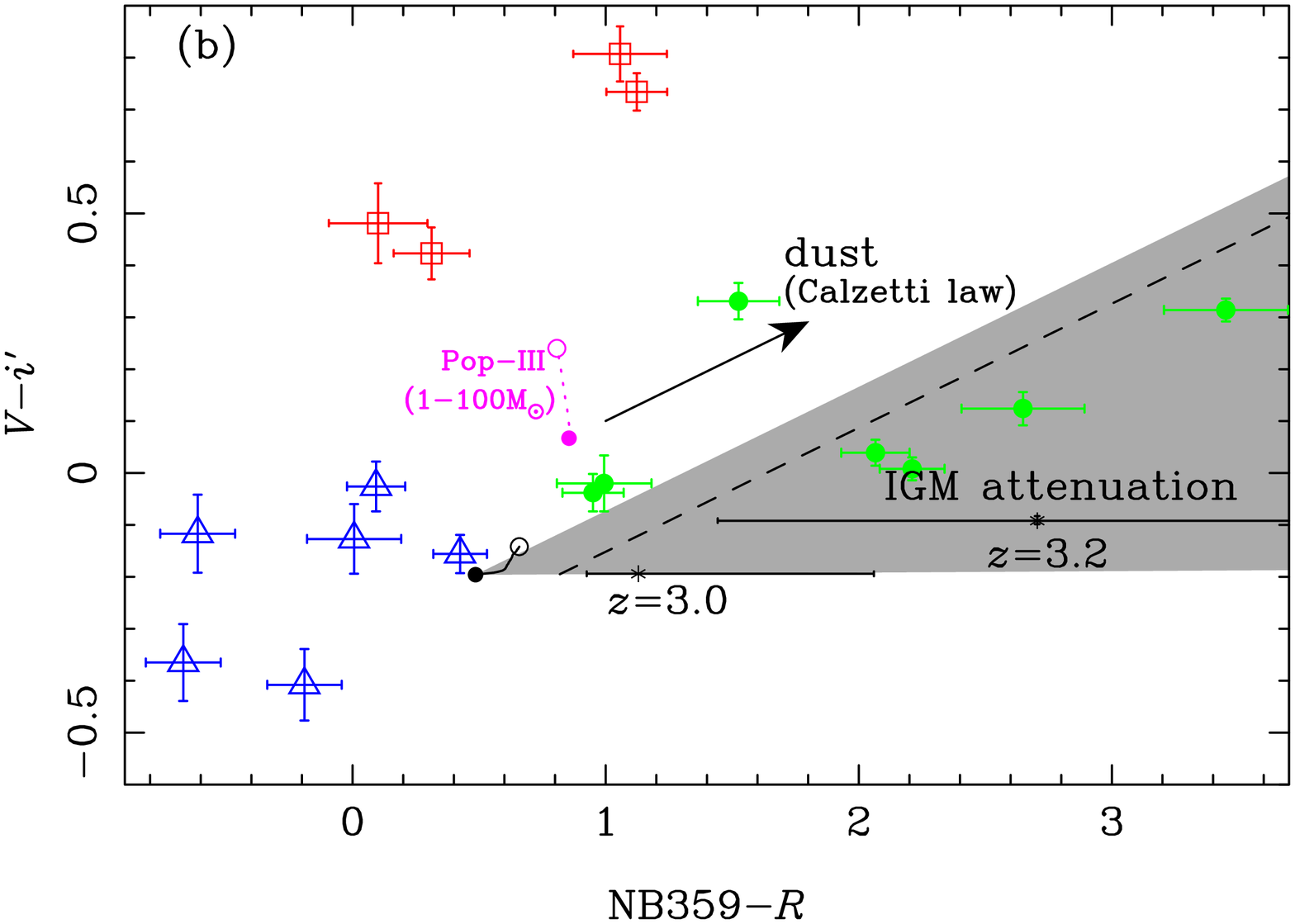}
\caption{(a) NB359$-R$ and $V-i'$ colors of objects detected in NB359.
 LBGs detected in NB359 are shown with circles, and two LAE sub-groups
 divided by their UV slopes are shown with open squares and 
 triangles. Filled and open circles connected with a olid line are the
 color tracks of a model galaxy with the bluest SED with the Salpeter
 IMF from $z=3.0$ and 3.3, respectively. 
 The shaded area indicates a color range which can be explained with
 attenuation by dust and IGM on this model SED. 
 The arrow indicates the direction of dust attenuation following a
 prescription by \citet{calzetti00} and changes in colors with
 $E(B-V)=0.1$ attenuation. Two filled circles with horizontal error bars
 show the colors of the bluest model SED with the median IGM attenuation
 at $z=3.0$ and $z=3.2$. The error bars represent the ranges of IGM
 opacity with 68\% probability. 
 The expected colors with the stellar population
 at zero metallicity based on the SED by \citet{schaerer03} from $z=3.0$
 to $z=3.3$ are also shown (labeled with ``Pop-III''). The model galaxy
 SEDs do not include nebular continuum emission. 
 (b) same as (a) for observed colors of the objects detected in NB359,
 but model galaxy SEDs include nebular continuum emission.
\label{fig:color2}}
\end{figure}

The {\it relative} escape fraction is defined as \citep{steidel01}:
\begin{equation}
 f_\mathrm{esc,rel} = \frac{(f_{1500}/f_{900})_\mathrm{int}}
 {(f_{1500}/f_{900})_\mathrm{obs}} \mathrm{exp}(\tau_\mathrm{IGM, 900}),
\end{equation}
where $(f_{1500}/f_{900})_\mathrm{int}$ and 
$(f_{1500}/f_{900})_\mathrm{obs}$ are the intrinsic and observed UV to 
Lyman continuum flux density ratios, respectively, and 
$\tau_\mathrm{IGM,900}$ is the IGM optical depth for Lyman continuum
photons along the line of sight. 
For the ease of comparison with previous studies on Lyman continuum at 
$z\sim3$ (Steidel et al. 2001; Inoue et al. 2005; S06), 
first we use $(f_{1500}/f_{900})_\mathrm{int}=3.0$.
If the dust attenuation at 1500 \AA\ $A_{1500}$ is known, 
$f_\mathrm{esc,rel}$ can be converted to $f_\mathrm{esc}$ as 
$f_\mathrm{esc} = 10^{-0.4A_{1500}} f_\mathrm{esc,rel}$
\citep{inoue05,siana07}.
We calculate the observed flux densities at rest-frame 1500 \AA\ by
interpolating flux densities in $V$ and $R$-bands, and use flux 
densities with the NB359 filter as $f_\mathrm{900,obs}$. 
For different LBGs, different aperture sizes are used to eoncompass the
bulk of fluxes and to avoid contamination from neighbours (see section
4.2).
For the seven detected LBGs, $(f_{1500}/f_{900})_\mathrm{obs}$ ranges
from 2.4 to 23.8 with a median value of 6.6.
If we consider the case without IGM attenuation, this median value
corresponds to $f_\mathrm{esc,rel}=0.45$, and it is a lower limit of
$f_\mathrm{esc,rel}$ (for $(f_{1500}/f_{900})_\mathrm{int}=3.0$).
Under the assumption of dust attenuation $E(B-V)=0.15$,
which would be reasonable to explain their $V-i'$ colors (see Figure
\ref{fig:color2}), a lower limit of $f_\mathrm{esc}$ is 0.11. 
If a correction for the IGM attenuation for a median opacity 
through the NB359 filter $\tau_\mathrm{IGM}=0.59$ (for $z=3.0$) is
applied, $f_\mathrm{esc}=0.20$. 
One unevitable obstacle to estimate escape fraction from observed flux
density ratios is the uncertainty of the intrinsic UV to Lyman continuum
flux density ratio which cannot be observed. 
For the bluest model galaxy generated with Starburst99 (top-heavy IMF,
low metallicity, no dust attenuation, no nebular continuum) considered
in section 4.3, $(f_{1500}/f_{900})_\mathrm{int}$ is 1.07. If we adopt
this intrinsic flux ratio instead of 3.0, estimated values of the escape 
fraction drop. For the case of no IGM attenuation and $E(B-V)=0.15$, 
$f_\mathrm{esc,rel}$ is 0.16 and $f_\mathrm{esc}$ is 0.04. 
In Table \ref{tab:uvtolc} we summarize our estimates of
$f_\mathrm{esc,rel}$ and $f_\mathrm{esc}$ for LBGs, as well as the
average values of $f_\mathrm{esc,rel}$ obtained by \citet{steidel01} and
S06.
Although there are uncertainties on  $(f_{1500}/f_{900})_\mathrm{int}$,
amount of dust attenuation and IGM opacity, the high
$f_\mathrm{esc,rel}$ for the LBGs at $z\sim3$ we obtained makes a clear
contrast with a stringent upper limits $f_\mathrm{esc,rel}<0.08$ for
stacked images of $z\sim 1.3$ galaxies \citep{siana07}. 
Also, five among the seven detected LBGs have
$(f_{1500}/f_{900})_\mathrm{obs}$ less than 10. Thus the UV-to-LC ratios
of the LBGs detected by us are smaller than the average values of LBGs
obtained by \citet{steidel01} and S06.
However, we should emphasize that, as it is seen in Figure \ref{fig:color1},
these LBGs should be the systems with smallest 
$(f_{1500}/f_{900})_\mathrm{obs}$ among luminous galaxies at 
$z \sim 3$. The average $f_\mathrm{esc}$ of the entire galaxy populations
(at least for $L \gtrsim L^\ast$ galaxies) should be much lower than the
value derived for our detected LBG sample. 
The constraints on $f_\mathrm{esc}$ for UV-magnitude limited sample with
the present data will be discussed in a forthcoming paper.
Also, we should remind that the SSA22 field is a prominent proto-cluster
at $z=3.09$, and the properties of ionizing radiation in such an
uncommon environment may not be representative of the epoch. A census
over a blank field would be required to investigate this point.


\acknowledgments

We thank Alex Razoumov for a careful read of the manuscript and
discussions. 
We would also like to thank the anonymous referee for helpful comments. 
II also thanks D. Schaerer for providing the SED data of his Pop-III
model, and M. Sawicki and J. P. U. Fynbo for stimulating discussions. 
II and AKI acknowledge Grant-in-Aid for Young Scientists (B; 18740114)
from Japan Society for the Promotion of Science (JSPS) 
and support by the Institute for Industrial Research, Osaka Sangyo
University.
We would like to express our acknowledgment to the indigenous Hawaiian
community for understanding of the significant role of the summit
of Mauna Kea in astronomical research.

\clearpage

\begin{deluxetable}{llcccc}
\tabletypesize{\footnotesize}
\tablecaption{Compilation of observed UV to LC flux ratios
 $(f_{1500}/f_{900})_\mathrm{obs}$ and escape fractions 
 of LBGs at $z \sim 3$. }
\tablehead{\colhead{References} & \colhead{Sample} &
 \colhead{$(f_{1500}/f_{900})_\mathrm{obs}$} &
 \colhead{$(f_{1500}/f_{900})_\mathrm{int}$\tablenotemark{a}} &
 \colhead{$f_\mathrm{esc,rel}$} &
 \colhead{$f_\mathrm{esc}$}}
\startdata
\citet{steidel01} & 29 LBGs, Average & 17.7$\pm$3.8 & 3.0 & 0.31 &  \\
\citet{shapley06} & 2 LBGs, Direct   & 12.7$\pm$1.8, 7.5$\pm$1.0 & 3.0 & 0.43,0.72 & \\
\citet{shapley06} & 14 LBGs, Average & 58$\pm$25 & 3.0 & 0.094 & \\
This work         & 7 LBGs, Direct   & 6.6 (median) & 3.0  & 0.46\tablenotemark{b} & 0.11\tablenotemark{c} \\
This work         & 7 LBGs, Direct   & 6.6 (median) & 3.0  & 0.83\tablenotemark{d} & 0.20\tablenotemark{c} \\
This work         & 7 LBGs, Direct   & 6.6 (median) & 1.07 & 0.16\tablenotemark{b} & 0.04\tablenotemark{c} \\
This work         & 7 LBGs, Direct   & 6.6 (median) & 1.07 & 0.30\tablenotemark{d} & 0.07\tablenotemark{c} \\
\enddata
\label{tab:uvtolc}
\tablenotetext{a}{$(f_{1500}/f_{900})_\mathrm{int}$ is an
 assumed intrinsic flux ratio used to derive $f_\mathrm{esc,rel}$ and
 $f_\mathrm{esc}$ from the observed UV to LC flux ratio. 
 $(f_{1500}/f_{900})_\mathrm{int}=3.0$ has been assumed in the previous
 studies shown here, and the case $(f_{1500}/f_{900})_\mathrm{int}=1.07$
 corresponds to the bluest model galaxy SED generated with the Starburst
 99 code.}
\tablenotetext{b}{no IGM attenuation is assumed.}
\tablenotetext{c}{$E(B-V)=0.15$ is assumed.}
\tablenotetext{d}{a median opacity at $z=3.0$ through the NB359 filter
 $\tau_\mathrm{IGM}=0.59$ is assumed.}
\end{deluxetable}

\clearpage 

\end{document}